\documentclass[runningheads]{llncs}
\usepackage{ragged2e}

\usepackage[hyphens]{url}
\usepackage{lipsum}
\usepackage{graphicx}
\usepackage{pdfpages}
\usepackage{enumitem}
\usepackage{hyperref}
\usepackage{blindtext}
\usepackage{dirtytalk}
\usepackage{hyperref}
\usepackage{amssymb}
\usepackage{amsmath}
\usepackage{colortbl} % For coloring
\usepackage{xcolor} % For text coloring
\usepackage{tcolorbox}

\usepackage[hyphens]{url}
\usepackage{lipsum}
\usepackage{graphicx}
\usepackage{pdfpages}
\usepackage{enumitem}
\usepackage{hyperref}
\usepackage{blindtext}
\usepackage{dirtytalk}
\usepackage{hyperref}
\hypersetup{ % Configure the package to make citations clickable
  colorlinks=true,
  linkcolor=black,
  citecolor=blue,
  urlcolor=blue
}

\usepackage{todonotes}

\usepackage[utf8]{inputenc}
\usepackage[T1]{fontenc}
\usepackage{newtxtext}
\usepackage[ngerman,italian,english]{babel}
\usepackage{csquotes}
\usepackage{amsmath, amssymb}
\usepackage{graphicx}
\usepackage[
    backend=biber,natbib,
    citestyle=ieee
]{biblatex} % Note that you must run ``biber'' to generate the bibliography
\begin{document}

\title{Predicting known Vulnerabilities from Attack News: A Transformer-Based Approach}

% \author{Anonymous Authors}
% \institute{Anonymous Institute}

\author{Refat Othman\inst{1} \and Diaeddin Rimawi\inst{1} \and Bruno Rossi\inst{2} \and 
Barbara Russo\inst{1}}
\institute{Free University of Bozen-Bolzano, Bolzano, Italy \\
\email{\{rothman, diaeddin.rimawi, barbara.russo\}@unibz.it} \and Masaryk University, Brno, Czech Republic  \\ \email{\ brossi@mail.muni.cz}}

\maketitle

\begin{abstract}
Identifying the vulnerabilities exploited during cyberattacks is essential for enabling timely responses and effective mitigation in software security. This paper directly examines the process of predicting software vulnerabilities, specifically Common Vulnerabilities and Exposures (CVEs), from unstructured descriptions of attacks reported in cybersecurity news articles. We propose a semantic similarity-based approach utilizing the \texttt{multi-qa-mpnet-base-dot-v1} (MPNet) sentence transformer model to generate a ranked list of the most likely CVEs corresponding to each news report.
To assess the accuracy of the predicted vulnerabilities, we implement four complementary validation methods: filtering predictions based on similarity thresholds, conducting manual validation, performing semantic comparisons with the first vulnerability explicitly mentioned in each report, and comparing against all CVEs referenced within the report. Experimental results, drawn from a dataset of 100 SecurityWeek news articles, demonstrate that the model attains a precision of 81\% when employing threshold-based filtering. Manual evaluations report that 70\% of the predictions are relevant, while comparisons with the initially mentioned CVEs reveal agreement rates of 80\% with the first listed vulnerability and 78\% across all referenced CVEs. In 57\% of the news reports analyzed, at least one predicted vulnerability precisely matched a CVE-ID mentioned in the article. These findings underscore the model’s potential to facilitate automated vulnerability identification from real-world cyberattack news reports.
\end{abstract}

\section{Introduction}\label{sec:introduction}
Timely response to cyberattacks is essential to limit damage and prevent further exploitation. News reports frequently provide early details about such incidents, often before technical reports are available. When a cyberattack is reported in the news, it is crucial to quickly identify and link the underlying vulnerabilities exploited. Delays in establishing this connection can leave systems exposed to ongoing threats, increasing both risk and remediation costs. Cybersecurity threats have become increasingly pervasive, with organizations experiencing an average of over 1,000 attacks per week~\cite{CResearch}. Check Point Research reported a 28\% increase in the frequency of such attacks during the first quarter of 2024 compared to the prior quarter~\cite{2024Report}. This escalation can be attributed to the rapid advancement of technology, which has enabled cybercriminals to devise innovative methods for exploiting system vulnerabilities; in fact, the number of identified vulnerabilities has surged by 25\% over the past two years~\cite{NVDvulnerability}.
In response to these growing threats, various initiatives have been developed to enhance organizational defenses, among which Cyber Threat Intelligence (CTI) has emerged as a critical strategy~\cite{liu2022context2vector}. CTI encompasses the systematic process of collecting, analyzing, and disseminating information concerning potential cyber threats and vulnerabilities that may compromise an organization’s security posture~\cite{CTI, CTIINfo}. 

A prominent resource utilized within CTI is the MITRE family of repositories~\cite{ATTACK}, which provides publicly available information regarding contemporary attacks and software/hardware vulnerabilities. This family comprises several key components: (1) The Adversarial Tactics, Techniques, and Common Knowledge (ATT\&CK) repository~\cite{ATTACK}, which catalogues various attack tactics, techniques, and procedures. (2) Common Vulnerabilities and Exposures (CVE)~\cite{CVEdataset}, which details known vulnerabilities\footnote{A \textit{vulnerability} is defined as a flaw in software code that has been exploited, potentially impacting the availability, confidentiality, or integrity of an organization’s assets~\cite{elder2022really, othman2024cybersecurity}} along with the affected systems and products. (3) The Common Weakness Enumeration (CWE)~\cite{CWE}, a collaboratively developed catalog of software weaknesses, coding errors, and security flaws. (4) The Common Attack Pattern Enumeration and Classification (CAPEC)~\cite{CAPEC}, which offers a compilation of recognized attack patterns that exploit known weaknesses in systems and products~\cite{lonetti2023model, othman2024comparison,othman2024vulnerability}.

These repositories provide a remarkable set of information for CTI. However, navigating and linking such a large amount of information poses challenges. This work aims to contribute to CTI  by providing a methodology and an approach to automatically link attacks to vulnerabilities exposed by software systems. Specifically, we offer cybersecurity researchers and practitioners a CTI model derived from the textual information of the MITRE repositories that predicts known vulnerability descriptions from reported attacks.
Linking attack-related news reports to specific CVEs can assist the cybersecurity community in understanding emerging threats and response priorities. 
To illustrate the need for such automation, consider a security analyst who is reviewing a news report describing a newly discovered breach. At this early stage, information vulnerability such as CVEs is often not available, yet rapid insights are essential for prioritizing response actions. Our approach addresses this gap, as soon as an article is published, the system can automatically suggest which known vulnerability is most likely being exploited, enabling analysts to quickly assess exposure, initiate patching, and guide incident response decisions. 
% This functionality is intended to support threat intelligence analysts, incident responders, and security operations teams by integrating seamlessly into existing CTI workflows. 
For instance, the model can be embedded within CTI platforms or news monitoring dashboards to automatically flag relevant CVEs whenever new attack reports appear, reducing the time and manual effort required to link emerging threats with known vulnerabilities.
However, linking attacks manually with 295,604 CVE issues~\cite{CVEdataset} is a non-trivial task requiring automated models to link attack information with vulnerabilities and weaknesses. 
% #its fine-tuning on large-scale question–answer datasets optimizes it for semantic similarity
% tasks, enabling it to capture nuanced contextual dependencies and semantic relations
% within short, technical texts that are particularly beneficial for vulnerability inference

This paper presents a novel method for automating vulnerability detection in cyberattack news articles using the best-performing sentence-transformer identified in our prior studies~\cite{othman2025attack,othman2024cybersecurity}, \texttt{multi-qa-mpnet-base-dot-v1 (MPNet)}.
Among all models evaluated in our earlier work~\cite{othman2025attack,othman2024cybersecurity}, MPNet achieved the highest accuracy due to its hybrid pre-training scheme, which combines masked and permuted language modeling. This training strategy equips MPNet with the ability to capture long-range dependencies and subtle semantic links between attack narratives and vulnerability descriptions. Moreover, its fine-tuning on large-scale question–answer datasets further enhances its semantic similarity capabilities, making it particularly effective for analyzing short, technical security text.
While MPNet has previously shown strong performance in linking MITRE-curated attack descriptions with CVE records~\cite{othman2025attack}, its effectiveness on unstructured, real-world cybernews has not yet been investigated. To address this gap, we introduce the first dataset that annotates SecurityWeek articles~\cite{SecurityWeek} with their corresponding vulnerabilities, enabling systematic and reproducible evaluation in this domain.
To support this task, we adopt a transfer-learning strategy in which MPNet is first fine-tuned on structured MITRE sources, where attack and CVE descriptions provide clean, well aligned semantic signals, and subsequently adapted to the very different linguistic characteristics of cyberattack news. Thus, we also release an initial prototype that applies the transfer-learned, fine-tuned MPNet model to news articles, enabling automated estimation of the vulnerabilities likely implicated in reported incidents. This prototype demonstrates how transfer learning, leveraging a model originally optimized for structured MITRE attack descriptions, can be effectively adapted to analyse unstructured real-world cyberattack news.
Both our annotated dataset~\cite{VULDATDataSet} and the accompanying implementation for model inference, transfer-learning fine-tuning, and validation are publicly available on GitHub~\cite{VULDAT} to support reproducibility and future work.

Thus, we intend to answer the following research questions:
\par\noindent

\newcommand{\rqone}{\textbf{RQ1:} \textit{To what extent can our approach predict software vulnerabilities from free-form textual descriptions of cyberattacks news?}}
\newcommand{\rqtwo}{\textbf{RQ2:} \textit{How effective are our oracle-based validation methods in assessing the accuracy of CVE predictions from attack descriptions?}}

\begin{itemize}
    \item[]
    \rqone
    
    To answer this question, we conduct a manual evaluation of the CVEs predicted by our model using a dataset of 100 cybersecurity news reports from SecurityWeek. Each report is processed using a sentence transformer to generate a top-$K$ list of likely CVEs based on semantic similarity. We manually reviewed each predicted CVE by comparing its description to the content of the corresponding news report, determining whether the predicted vulnerability was relevant to the attack described. 
    % This manual validation resulted in a 70\% relevance rate across 2000 predictions, providing evidence that our approach is capable of identifying contextually appropriate CVEs from unstructured attack reports.
    \item[] \rqtwo

    To answer this question, we applied three automated validation methods based on cosine similarity. These methods respectively assess: (i) the semantic relevance of predicted CVEs above a similarity threshold, (ii) alignment with the first CVE mentioned in each report, and (iii) consistency with all CVEs cited in the news report.     
    % These results suggest that our oracle-based methods provide a strong and scalable alternative to manual inspection.
\end{itemize}
Overall, the major contributions of our work are the following:
\begin{itemize}
% no information , not yet information  vulnerability

% same semantic  different terminology

% test our model on free form textual, 
% test the our approach for 3 different methods 
% task transferlearning 
\item A new application of an MPNet-based sentence-transformer model to real-world cyberattack news data;
\item A multi-method evaluation framework combining expert manual analysis with three oracle-based validation techniques;
\item An empirical study on 100 SecurityWeek articles, evaluated through both manual expert review and automated validation;
\item A practical contribution to CTI automation by demonstrating how transformer-based semantic similarity can support early vulnerability identification from unstructured text;

\end{itemize}

The paper is structured as follows: Section~\ref{Sec:background} briefly summarises the background, key concepts, and prior studies on software vulnerabilities. Section~\ref{Sec:methodology} discusses our methodology, including the tool and dataset. Section~\ref{sec:result} shows the results of our approach. Section~\ref{sec:threatstovalidity} identifies the limitations and summarizes the threats to validity. Section 6 discusses the related work, and the paper concludes in Section~\ref{ch:conclusion}.

\section{Background}
\label{Sec:background}
In this section, we review the core concepts of our study: vulnerability knowledge bases, cyberattack news for threat intelligence, and the MPNet transformer model.
% we review the concepts underpinning our research: vulnerability knowledge bases and the use of cyberattack news as a source of threat intelligence. It also covers the application of natural language processing techniques and the MPNet sentence transformer model, which plays a significant role in our current study.

\subsection{Vulnerability Knowledge Bases}
\label{sec:vulnerability}
% \begin{figure}[htb!]
% \centering
% % \begin{subfigure}[b]{0.4\textwidth}
% % \includegraphics[scale=0.6]{Fig/Relationship MITRE ATTACK.drawio.pdf}
% % \caption{ATT\&CK model.\label{fig:attackModel}}
% % \end{subfigure}\hspace{8pt}
% % \begin{subfigure}[b]{0.6\textwidth}
% \includegraphics[scale=0.5]{Fig/CVEExample.pdf}
% \caption{An example of CVE-2025-4754.}
% \label{fig:CVEexample}
% % \end{subfigure}
% % \caption{The attack model.}\label{sec:attackmodel}
% \end{figure}
According to the NIST National Vulnerability Database~\cite{NVDvulnerability}, a vulnerability is a flaw in the computational logic of software or hardware that, if exploited, can compromise confidentiality, integrity, or availability. Such vulnerabilities are core to the cybersecurity threat landscape, as they are commonly exploited to gain unauthorized access, execute malicious code, or disrupt services~\cite{elder2022really, WhatCVE, alevizopoulou2021social, gasmi2019information}. Given the unpredictability of when and how a vulnerability will be exploited, it is essential to equip stakeholders with effective tools to detect and mitigate them~\cite{iorga2021yggdrasil, queiroz2019eavesdropping, baccar2021automated, dionisio2019cyberthreat, tang2023csgvd}.
The CVE repository provides a centralized catalog of publicly disclosed software vulnerabilities~\cite{WhatCVE, catillo2021critique,regano2024privacy}, assigning a unique CVE identifier to each entry for consistent cross-referencing across systems. Each CVE typically includes a brief description, identification number, and external references. Many entries also reference CWE categories~\cite{CWE}, which classify underlying weakness types. Additional metadata may include patch details, severity scores based on the Common Vulnerability Scoring System (CVSS)~\cite{CVSS}, and impact assessments. 
Many vulnerabilities remain undetected until exploited, as traditional methods struggle with growing threat complexity. Analyzing unstructured sources like cyber news offers a proactive path to improve defense and reduce mitigation efforts~\cite{vuldatPaper222}.
% Despite the availability of such structured data, vulnerabilities often remain undiscovered until they are exploited in real-world attacks. The growing volume and sophistication of cyber threats challenge traditional detection methods, prompting a shift toward more proactive and intelligent approaches. Early identification of vulnerabilities through analysis of attack indicators, especially in unstructured sources such as cyber news reports, offers a promising direction to improve system defenses and reduce mitigation efforts.

\subsection{Cyberattack News}
\label{sec:attackNews}
Cyberattack reports in the form of news reports, incident disclosures, blogs, and advisories have emerged as rich sources of real-time threat intelligence. These texts typically describe adversarial behavior, targeted systems, exploited vectors, and observed outcomes. Unlike structured databases, news-based reports provide contextual and chronological information that may reveal unknown or unlinked vulnerabilities. As such, they represent an underutilized yet valuable asset for proactive security analysis~\cite{husari2017ttpdrill}.

Extracting actionable intelligence from these texts, however, is non-trivial. The descriptions are often ambiguous, vary in terminology, and lack explicit references to known vulnerabilities~\cite{wang2024knowcti}. For example, a news report may mention that attackers gained access by exploiting a weakness in a widely used library without naming the corresponding CVE. This lack of specificity significantly limits the effectiveness of automated extraction systems, particularly those dependent on strict keyword matching or predefined rule-based approaches.
To address this challenge, natural language processing (NLP) techniques extract meaningful information from unstructured text~\cite{jo2022vulcan}. 
Combining NLP methods with structured cybersecurity repositories makes it possible to uncover links between textual attack descriptions and known vulnerabilities even without explicit identifiers. Moreover, the effectiveness of these methods is strongly influenced by the model's capability to understand complex semantic relationships within the text.

\subsection{NLP and MPNet Sentence Transformer}\label{sec:transformer}
Sentence transformer models are pre-trained models that produce sentence embeddings for various natural language processing tasks, such as semantic search, paraphrasing, and clustering. However, these models usually contain hundreds of millions of parameters, which brings challenges for fine-tuning and online serving in real-life applications for latency and capacity constraints.
In this work, we use the \texttt{multi-qa-mpnet-base-dot-v1} model~\cite{multi-qa-mpnet-base-dot-v1}, which is based on the MPNet architecture~\cite{song2020mpnet}. 
MPNet is an improved Transformer-based model that extends BERT~\cite{devlin2018bert} by integrating both masked language modeling and permuted language modeling, allowing the model to capture dependencies between tokens better and generate richer semantic embeddings. 
The \texttt{multi-qa-mpnet-base-dot-v1} model has been fine-tuned on over 200 million question-answer pairs from diverse domains~\cite{multi-qa-mpnet-base-dot-v1}, making it particularly effective for semantic search and question-answer retrieval. The MPNet model outputs 768-dimensional vectors and can encode a sentence's meaning with high precision. In this study, we use pre-trained transformer models to generate embeddings for attack and vulnerability descriptions and then apply a similarity layer based on cosine similarity to determine whether an attack is linked to a vulnerability (Section~\ref{sec:Arch}).

% Performance Evaluation Metrics
\section{Methodology}
\label{Sec:methodology}

% \begin{figure}[htp!]
% \centering
% \includegraphics[width=.7\columnwidth]{Fig/VULDATNewOverview.drawio.pdf}
% \caption{VULDAN pipeline.}
% \label{fig:overview}
% \end{figure}

\begin{figure}[hbt!]
\centerline{\includegraphics[width=.95\columnwidth]{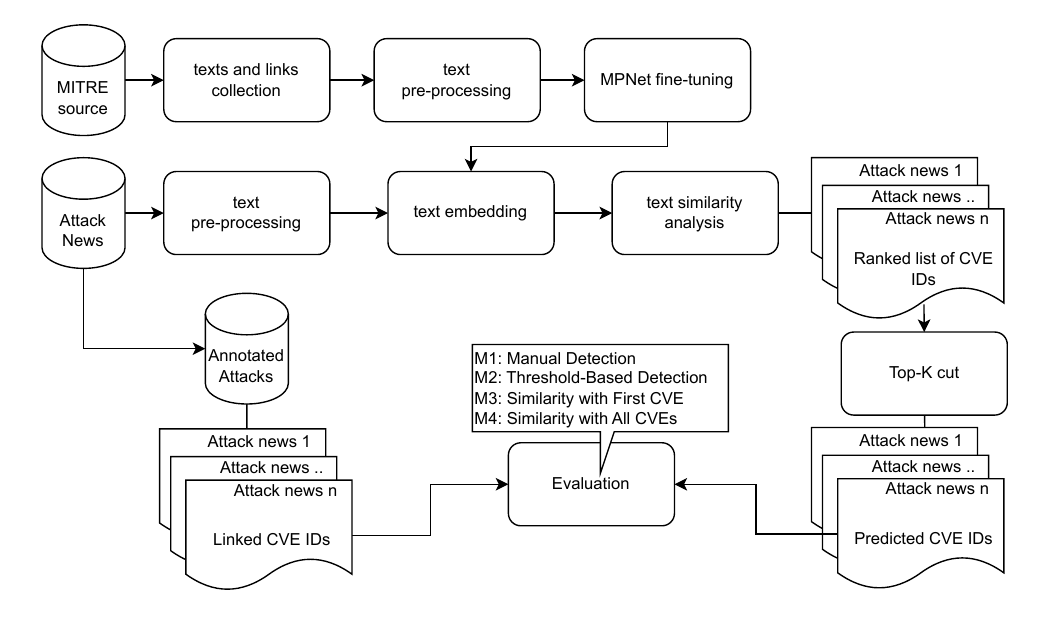}}
\caption{Overview of methodology.}
\label{fig:methodology}
\end{figure}

In this section, we describe our research methodology for detecting software vulnerabilities based on real-world cyberattack news, as illustrated in Fig.~\ref{fig:methodology}. The process starts with the collection of a dataset of cybersecurity news reports~(Section~\ref{sec:mapprocess}). This is followed by a comprehensive text pre-processing step to clean and standardize the textual data~(Section~\ref{sec:textpreprocessing}). Next, we employ the fine-tuned MPNet Sentence Transformer to generate high-dimensional embeddings for both the processed news reports and the known CVE vulnerability descriptions~(Section~\ref{sec:Arch}). A similarity calculation is then performed between each news embedding and all CVE embeddings using cosine similarity, discussing sensitivity analysis~(Section~\ref{sec:sensitivityAnalysis}). Finally, the performance of the detection method is evaluated against a ground truth using the validation strategies M1-M4~(Section~\ref{sec:evaluation}).

\subsection{Data Collection}
\label{sec:mapprocess}

\par\noindent

\par\noindent
\textbf{\textit{MITRE source (Attacks-vulnerabilities Mapping, $\mathcal{M}$):}}
% To fine-tune our MPNet-based model for vulnerability prediction, we constructed a mapping $\mathcal{M}$ between textual descriptions of cyberattack techniques and known vulnerabilities. This mapping is derived from the MITRE repositories, primarily ATT\&CK, CAPEC, CWE, and CVE. 
% $$\mathcal{M}: A\rightarrow C \\$$
% $$\mathcal{M}(a) =\{ c \in \mathcal{C}   \, : \, \exists  \, a \rightarrow c\}$$
% where $C$ is the set of all CVE issues, $A$ is the set of all attack techniques reports in ATT\&CK.
% The mapping associates an attack text, $a$, to a list of  CVE issues through \textit{explicit links} ($\rightarrow$) derived from the information contained in the repositories ATT\&CK, CAPEC, CWE, and CVE. 
% We leveraged the mapping $\mathcal{M}$ to create a dataset of attack techniques and vulnerabilities annotated according to the links in $\mathcal{M}$. The annotated dataset~\cite{VULDATDataSetReview} is used as ground truth for fine-tuning of the model.
% \subsection{Data Collection (Attack--Vulnerability Mapping, $\mathcal{M}$)}
% \label{sec:mapprocess}
The MITRE-based dataset used to fine-tune our MPNet model for vulnerability prediction. Specifically, it provides the ground-truth associations between attack descriptions and the CVE reports.
In what follows, we describe the construction of the mapping $\mathcal{M}$ that links attack descriptions to CVE entries. These connections are drawn from the explicit relationships documented across the ATT\&CK, CAPEC, CWE, and CVE repositories.

\[
\mathcal{M}: A \rightarrow C
\]
\[
\mathcal{M}(a)=\{c\in\mathcal{C} \; | \; \exists\, a \rightarrow c\}
\]

Here, $A$ denotes the set of attack descriptions in ATT\&CK and CAPEC, and $C$ represents the set of all CVE records. Each attack description $a$ is mapped to a corresponding set of CVEs through the explicit cross-repository links ($\rightarrow$) documented in ATT\&CK, CAPEC, CWE, and CVE. To construct $\mathcal{M}$, we systematically reviewed the official repository pages and extracted every explicitly defined relationship.
% Figure~\ref{fig:graphM} visualizes the resulting network of repository entities and their interconnections. Nodes represent categories of attack- or vulnerability-related information, and the values inside the nodes indicate how many outgoing links lead to subsequent entities. Different colors distinguish the various data sources.
% % Moreover, there are no direct links from ATT\&CK Tactics, Techniques, or Procedures to entries in the CWE or CVE repositories. Links from Attack Patterns to techniques are unidirectional, while those between Techniques and Tactics, and between Attack Patterns and CWE entries, are bidirectional. CWE entries link only downward to CVE reports.

% \begin{table*}[htb!]
% \caption{Number of CVE and CWE reports linked and not linked to attack descriptions.}
% \label{tab:golddataset}
% \centering
% \bgroup
% \begin{tabular}{llccccc}
% \hline
%  & &\textbf{Tactic} &\textbf{Technique} &\textbf{Procedure} & \textbf{Attack Pattern} \\
% \hline
% \textbf{CWE reports} & linked& 117&79& 117& 117  \\
%  & not linked &818&856&818&818\\
% \hline
% \textbf{CVE reports} &linked&610&439& 610& 610 \\
% &not linked &294994&295165&294994 &294994 \\
% \hline
% \end{tabular}%
% \egroup
% \end{table*}

\begin{table*}[htb]
\caption{Attack descriptions linked and not linked to CVE reports.}
\label{tab:golddataset2}
\centering
\bgroup
\setlength{\tabcolsep}{4pt}
\begin{tabular}{lccccccc}
\hline  
& \textbf{Tactic} &\textbf{Technique} &\textbf{Procedure} & \textbf{Attack Pattern} \\
\hline
 \textbf{Linked}&11 & 100  & 721 &86\\
\textbf{Not linked}&3&525&88&473\\
\hline
\textbf{Total}&14&625&809&559\\
\hline
\end{tabular}%
\egroup
\end{table*}

% In contrast to floating entries, certain nodes act as “super entries,” exhibiting a large number of incoming or outgoing links. These reflect areas where the relationships among adversarial behaviors, weaknesses, and vulnerabilities are more comprehensively documented. We propagate associations from subtechniques to CVE entries whenever a CAPEC pattern linked to the subtechnique is also explicitly connected to a CVE. A similar propagation is applied to Tactics based on their explicit links to Techniques.

Tables~\ref{tab:golddataset2} summarize the collected items and their link coverage across the different repository layers. In addition, the table shows the number of attack descriptions linked to CVE reports; most Tactics and Procedures contain such links, whereas the majority of techniques and Attack Patterns do not.
An example of the mapping structure is shown in Table~\ref{tab:cveDataset}, illustrating how CAPEC-38 relates to the ATT\&CK Technique T1574 and to multiple CVE entries, such as CVE-2022-4826 and CVE-2020-26284.

\begin{table*}[htbp!]
\centering
\caption{Example of a CAPEC pattern and its associated technique and CVE reports.}
\label{tab:cveDataset}
\begin{small}
\begin{tabular*}{\textwidth}{@{\extracolsep{\fill}}p{4cm}p{4cm}p{4cm}}
\hline
\textbf{CAPEC ID: Description} &\textbf{Technique ID: Description} &\textbf{CVE ID: Description} \\
\hline
\textbf{CAPEC-38:} This pattern describes how an adversary inserts a malicious component into a program’s trusted execution path so that the malicious code is executed instead of the legitimate resource. The attack may involve altering search paths or manipulating dependent libraries.
&
\textbf{T1574.007:} Adversaries may execute their own malicious payloads by hijacking environment variables used to load libraries. The PATH environment variable contains a list of directories (User and System) that the OS searches sequentially through in search of the binary that was called from a script or the command line.
&
\textbf{CVE-2022-4826:} A stored XSS vulnerability in the Simple Tooltips WordPress plugin before version 2.1.4 due to improper handling of shortcode attributes.
\\
&&
\textbf{CVE-2020-26284:} A command execution vulnerability in Hugo for Windows caused by loading malicious files from the working directory via PATH hijacking.
\\
\hline
\end{tabular*}
\end{small}
\end{table*}

We use the final mapping $\mathcal{M}$ to construct an annotated dataset linking attack descriptions to their associated weaknesses and vulnerabilities. This dataset~\cite{VULDATDataSet} serves as the ground truth for fine-tuning our MPNet model and for evaluating the accuracy of our vulnerability prediction approach.

% \begin{figure}[htp!]
% \centerline{\includegraphics[width=0.75\columnwidth]{Fig/textPreprocessing.drawio2.pdf}}
% \caption{Overview of the text pre-processing pipeline.}
% \label{fig:textpreprocessing}
% \end{figure}

\par\noindent
\textbf{\textit{SecurityWeek News Dataset:}}
% In this paper, we collected a dataset of the latest 100 cybersecurity news reports from SecurityWeek's~\cite{SecurityWeek} Vulnerabilities category, an online magazine and website that delivers news and expert analysis on cybersecurity threats, incidents, and vulnerabilities.
% These reports were selected to ensure diversity in attack types, affected systems, and reporting styles, enabling a broad evaluation of the model’s prediction capabilities.
% Moreover, it includes both news reports that explicitly reference known vulnerabilities (e.g., CVE-IDs) and those that describe technical details without directly naming specific CVEs. Out of the 100 reports, 97 include explicit mentions of CVE-IDs, while the remaining 3 do not reference any CVE in the text. This curated collection serves as a real-world benchmark to evaluate the effectiveness of our model in predicting vulnerabilities from unstructured attack text.
% \textbf{\textit{SecurityWeek News Dataset.}}
Following the construction of the MITRE-based fine-tuning dataset, we assembled a second, real-world evaluation dataset consisting of 100 cybersecurity news articles obtained from the \textit{Vulnerabilities} section of SecurityWeek~\cite{SecurityWeek}. SecurityWeek is a reputable cybersecurity news outlet known for reporting newly discovered vulnerabilities, active exploitation campaigns, emergency patches, and vendor advisories. 
The articles were collected chronologically to reflect the latest publicly disclosed incidents. 
Each article typically contains a mixture of high-level narrative, technical descriptions of the vulnerability or exploit behavior, details about affected software or vendors, references to exploitation status in the wild. We collected the latest 100 reports as they appeared on the website and subsequently extracted any CVE identifiers mentioned within them. For each article, we retrieved the full text along with title and applied pattern-based extraction to identify CVE references, followed by manual verification. We also recorded the vendors and products mentioned in the text using keyword extraction and manual confirmation.
Out of the 100 articles, 97 explicitly reference at least one CVE ID, while the remaining 3 discuss security flaws without naming specific vulnerabilities, providing useful test cases for implicit prediction scenarios. 
The dataset also captures a wide range of attack categories (e.g., remote code execution, privilege escalation, command injection, authentication bypass). This diversity ensures that the dataset covers variations in writing style, technical depth, and vulnerability classes, offering a realistic and challenging benchmark for evaluating the model’s ability to infer relevant CVEs from unstructured natural-language descriptions. The manually verified CVE annotations, together with the raw article content, form the basis for assessing the real-world applicability and generalization capability of our vulnerability prediction approach.

% \begin{table*}[htb]
% \centering
% \caption{Distribution of CVE mentions in attack news dataset.}
% \label{tab:attackNewsCVE2}
% \setlength{\tabcolsep}{6pt} % Adjust column spacing
% \renewcommand{\arraystretch}{1.2} % Adjust row spacing
% \resizebox{0.8\textwidth}{!}{ % Resize to fit within page width
% \begin{tabular}{lccc}
% \hline
% \textbf{}  & \textbf{\# of News reports mentioning CVE-IDs} & \textbf{\# of News reports without CVE mentions} \\ 
% \hline
% \textbf{Count} & 97 & 3 \\ 
% \hline
% \end{tabular}
% }
% \end{table*}

\subsection{Text pre-processing}
\label{sec:textpreprocessing}
We applied a light pre-processing pipeline designed to eliminate extraneous content while keeping the core semantics intact. We first standardized all text to lowercase and removed citations, embedded URLs, and other non-informative symbols. We intentionally avoided conventional NLP transformations such as stemming, lemmatization, and stop-word elimination. Prior research~\cite{okonkwo2023leveraging, siino2024text} indicates that transformer models operate on subword tokenization and derive much of their strength from contextual and syntactic information, which can be distorted by these techniques. Function words often provide structural cues that guide attention mechanisms, and morphological variations may carry subtle meaning differences that transformers can exploit. By applying only minimal cleaning while preserving linguistic and grammatical features, we ensure that the sentence transformer receives rich, contextually consistent input suitable for modeling relationships between attack descriptions and CVE entries.

\subsection{Approach Architecture}
\label{sec:Arch}

We implemented our approach using a sentence transformer model, augmented with a semantic similarity computation layer. Specifically, the model calculates a similarity score (Equation~\ref{eqSim}) between a given attack description and each CVE issue. For this work, we fine-tuned a single sentence transformer model, \texttt{multi-qa-mpnet-base-dot-v1}, which was identified in our previous study as the strongest-performing architecture for cybersecurity text matching~\cite{othman2025attack}. The model produces fixed-size sentence embeddings by encoding input texts through multiple transformer layers followed by mean pooling. Fine-tuning was performed on the ATT\&CK-to-CVE mapping dataset to encourage high cosine similarity between embeddings of linked attack–CVE pairs and low similarity for non-linked pairs. We used an 80/10/10 split for training, validation, and testing, respectively, following recommendations for reproducibility and stable comparison~\cite{HinidumaEtAl2025}. CosineSimilarityLoss was employed during training, with four epochs, 100 warmup steps, and evaluation every 500 steps, consistent with established practices in sentence-transformer training~\cite{hyperparams}. This fine-tuning process adapts MPNet to the specific linguistic and semantic characteristics of cybersecurity attack descriptions and vulnerability reports.

Following standard practice in embedding-based NLP~\cite{reimers2019sentence,muennighoff2022mteb}, we use cosine similarity as the standard metric for comparing sentence embeddings. Prior work has shown its effectiveness for semantic matching tasks: Reimers and Gurevych~\cite{reimers2019sentence} demonstrated strong gains in Sentence-BERT through cosine-based retrieval, while Muennighoff et~al.~\cite{muennighoff2022mteb} employed it as the default metric across more than 50 models in the MTEB benchmark. In our setting, cosine similarity (Equation.~\ref{eqSim}) is computed between the normalized embedding of an attack description $p$ and each CVE embedding $q$, producing values in the range $[-1,1]$ (practically $[0,1]$ for our normalized vectors). For consistency with our threshold analysis, these values are reported on a 0–100 scale. We then rank all CVE embeddings according to their similarity to the attack vector and retain those above a threshold $\rho$:
\[
\mathcal{L}_{\rho}(a)=\{c \in C \mid Sim(\vec{a},\vec{c}) > \rho\}.
\]
Here, $C$ denotes the set of all CVE IDs, and $\mathcal{L}_{\rho}(a)$ represents the predicted vulnerabilities for attack $a$. A classification decision is positive when $\mathcal{L}_{\rho}(a)$ is non-empty and negative otherwise.

After fine-tuning, we used the model to encode both attack descriptions and CVE issues into a shared semantic embedding space (see Section~\ref{sec:transformer}). Cosine similarity is then applied to the normalized embedding vectors to quantify their semantic closeness. The resulting similarity score ranges from 0 (no similarity) to 1 (maximum similarity):

\begin{equation}
\label{eqSim}
\text{Sim}(\vec{p}, \vec{q}) = \frac{\vec{p} \cdot \vec{q}}{|\vec{p}| \cdot |\vec{q}|} = \frac{\sum_{i=1}^{n} p_i q_i}{\sqrt{\sum_{i=1}^{n} p_i^2} \cdot \sqrt{\sum_{i=1}^{n} q_i^2}}
\end{equation}

Given an input attack text $a$, the model computes similarity scores between $a$ and all CVE descriptions. The CVEs are then ranked in descending order of similarity, forming a list $\mathcal{L}$. We evaluate the model’s performance based on varying values of $k$, where $k$ denotes the number of top-ranked CVEs considered. Accordingly, our approach links each attack description with the top-$k$ most semantically similar vulnerability reports.
All experiments were executed on a GPU cluster consisting of six nodes, each equipped with an NVIDIA A100 GPU (80\,GB), 192\,GB of system memory, and a 16-core Intel Xeon 4208 processor.

\subsection{Evaluation metrics and threshold/top-$k$ sensitivity analysis}\label{sec:sensitivityAnalysis}

To evaluate our model, we leveraged the ground truth mappings between attack texts and CVE issues derived from MITRE repositories. These mappings served for sensitivity analysis to determine the optimal value of $k$, which controls how many top-ranked CVEs are associated with each attack description. Specifically, for $a$ given attack text $a$, our model outputs a ranked list $\mathcal{L}(\textit{a})$ of CVE issues based on cosine similarity. We assess how well this list overlaps with the true set of CVEs $\mathcal{M}(\textit{a})$ from MITRE, where a correct prediction satisfies 

$$\mathcal{L}(\textit{a})\bigcap \mathcal{M}(\textit{a})\not = \emptyset$$

We compute the performance metrics by means of the cardinality of the sets in Table~\ref{tab:postivesNegatives}.
We consider each attack as an instance, where the model predicts a positive if $\mathcal{L}(\textit{a})$ is non-empty and a negative otherwise. Using this setup, we calculate standard classification metrics: Precision, Recall, and their harmonic mean (F1 score). These are computed as follows:
\par\noindent
% \textbf{Precision.}
%Precision is a metric for how well a classifier performs in recognizing instances of positive.  
% It is defined by:
\begin{equation}
\textit{Precision} = \frac{\textit{TP}}{\textit{TP} + \textit{FP}}
\end{equation}
%Precision measures the portion of correctly predicted CVE reports that are relevant.
\par\noindent
% \textbf{Recall.}
% Recall is defined by: 
\begin{equation}
% \begin{align*}
\textit{Recall} = \frac{\textit{TP}}{\textit{TP} + \textit{FN}}
% \end{align*}
\end{equation}
\par\noindent
% \textbf{F1.}
% \begin{align*}
\begin{equation}
\textit{F1} = 2 \times \frac{\textit{Precision} \times \textit{Recall}}{\textit{Precision} + \textit{Recall}}
% \end{align*}
\end{equation}

\begin{table}[htbp!]
\caption{Positives and Negatives.}
\label{tab:postivesNegatives}
\begin{center}
% Adjusting table width to 0.5\textwidth
\begin{tabular*}{0.6\textwidth}{@{\extracolsep{\fill}}p{0.22\textwidth}p{0.4\textwidth}}
\hline

% \begin{table}[htbp!]%
% % \scriptsize
% \caption{Positives and Negatives.}%
% \label{tab:postivesNegatives}
% \begin{center}
% \begin{tabular*}{0.5\textwidth}{@{\extracolsep{\fill}}ll}
\hline
\textbf{Type}  & \textbf{Description}  \\
\hline

% Positives & $\{ a \in \mathcal{A} \, : \exists \, {c} \in (\mathcal{C} \cap  \mathcal{M}_{a} )\}$\\
Positives & $\{ a \in \mathcal{A} \, : \exists \, {c} \in \mathcal{M}_{a} \}$\\
Negatives &  $\{ a \in \mathcal{A} \, : \nexists \, {c} \in \mathcal{M}_{a} \}$\\
% True Positive ($a$) & $\left| \mathcal{L}_{a} \cap \mathcal{M}_{a} \right| > 0$ \\
True Positives  (TP)& $\{ a \in \mathcal{A} \, :  \mathcal{L}_{a} \cap \mathcal{M}_{a} \not = \emptyset   \} $\\
% False Positive ($a$) &  $\left| \mathcal{L}_{a} - \mathcal{M}_{a} \right| > 0$\\
False Positives  (FP) & $\{ a \in \mathcal{A} \, :\left( \mathcal{L}_{a} \not = \emptyset   \, \land \, \mathcal{M}_{a}  = \emptyset  \right)  \}$ \\
False Negatives (FN)  & $\{ a \in \mathcal{A} \, : \left( \mathcal{L}_{a}  = \emptyset    \, \land \,  \mathcal{M}_{a}  \not = \emptyset   \right)\}$ \\
True Negatives (TN)  & $\{ a \in \mathcal{A} \, : \left( \mathcal{L}_{a}   = \emptyset   \,\land \,  \mathcal{M}_{a}  = \emptyset   \right)\}$ \\

\hline
\end{tabular*}
\end{center}
\end{table}

\begin{figure*}[htb!]
    \centering
    \includegraphics[width=0.8\textwidth]{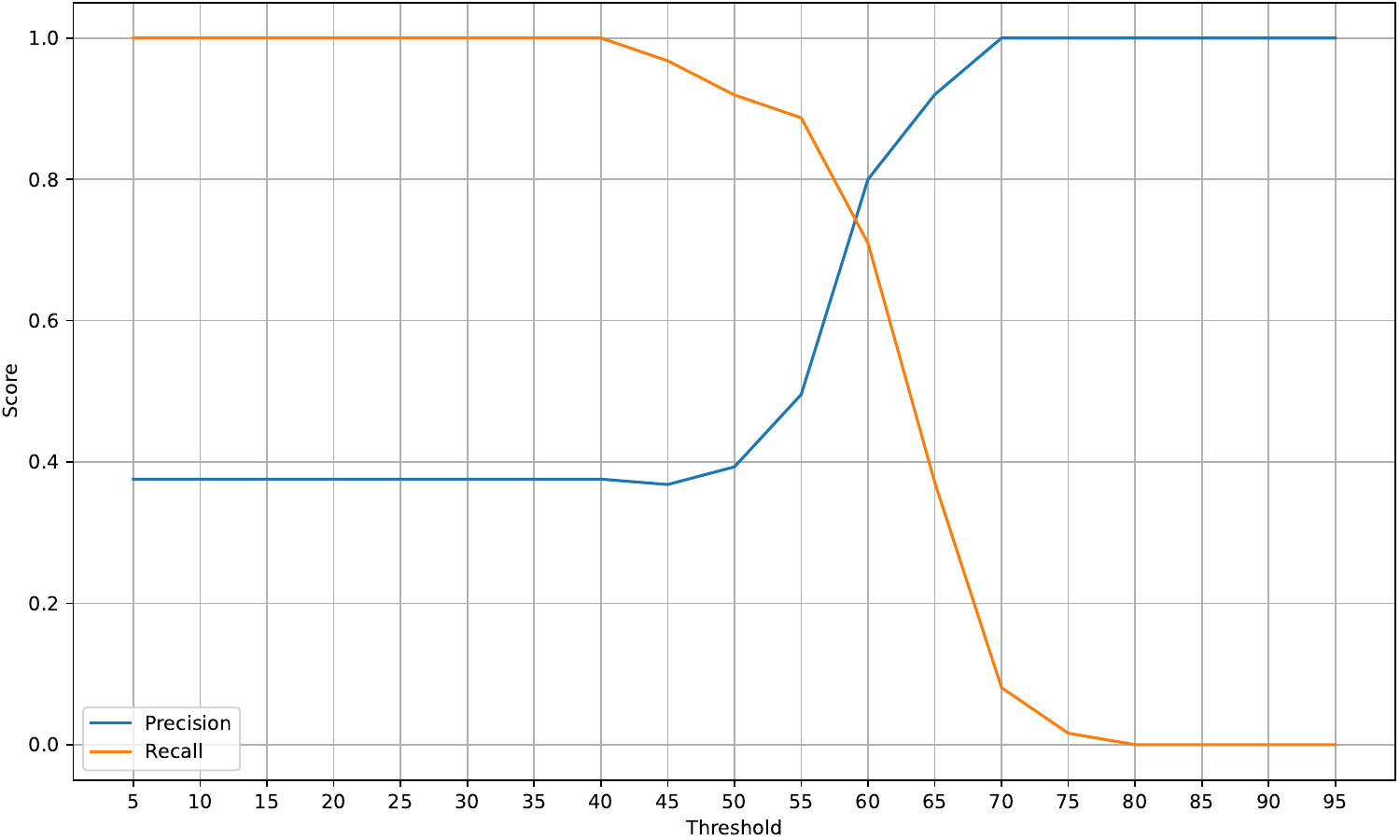}
    \caption{Precision-Recall curve for varying similarity threshold $\rho$.}
    \label{fig:sensitivityAnalysis}
\end{figure*}

In similarity-based classification tasks, a decision threshold $\rho$ is typically used to determine whether a predicted CVE is considered relevant to a given attack description. Although a default cutoff of $0.5$ is commonly adopted, prior work has shown that such arbitrary thresholds often fail to yield optimal performance~\cite{lobo2022cost,sheng2006thresholding}. To address this issue, we conducted a threshold-sensitivity analysis to identify the most effective value of $\rho$ for our model. Similar to standard practices in binary decision calibration, we examined how Precision and Recall vary across different threshold values using the Precision–Recall (PR) curve~\cite{davis2006relationship}. High precision reflects a low false-positive rate, whereas high recall indicates fewer false negatives. Using the \texttt{multi-qa-mpnet-base-dot-v1} model and a balanced subset of MITRE technique descriptions (59 positive and 59 negative pairs), we plotted the PR curve across a range of similarity thresholds. As depicted in Fig.~\ref{fig:sensitivityAnalysis}, the precision and recall curves intersect at $\rho = 0.58$, corresponding to the Equal Error Rate (EER) point. This threshold represents the most balanced trade-off between false positives and false negatives and is therefore adopted in our subsequent evaluations.

% Beyond the decision threshold, the choice of the top-$k$ predictions also plays a crucial role in ranking-based vulnerability identification. Selecting a small value of $k$ reduces false positives but risks missing relevant CVEs (i.e., higher false negatives), whereas larger values of $k$ increase coverage at the cost of introducing more irrelevant candidates. To determine an appropriate value of $k$, we carried out a sensitivity analysis using a balanced subset of ATT\&CK technique descriptions (50 positive and 50 negative instances). For each candidate $k$ value, we computed Precision and Recall scores and summarized their distributions using boxplots. Figure~\ref{fig:sensitivityAnalysisK} illustrates how these metrics evolve as $k$ increases. We observe that at $k = 20$, the average Precision and Recall curves converge, yielding the highest F1 score and indicating the best balance between completeness and correctness. Accordingly, we set $k = 20$ in our main experiments, as this value provides the most reliable selection of relevant CVEs for each attack description.

\begin{figure}[htb]
\centering
\small
\includegraphics[width=\columnwidth]{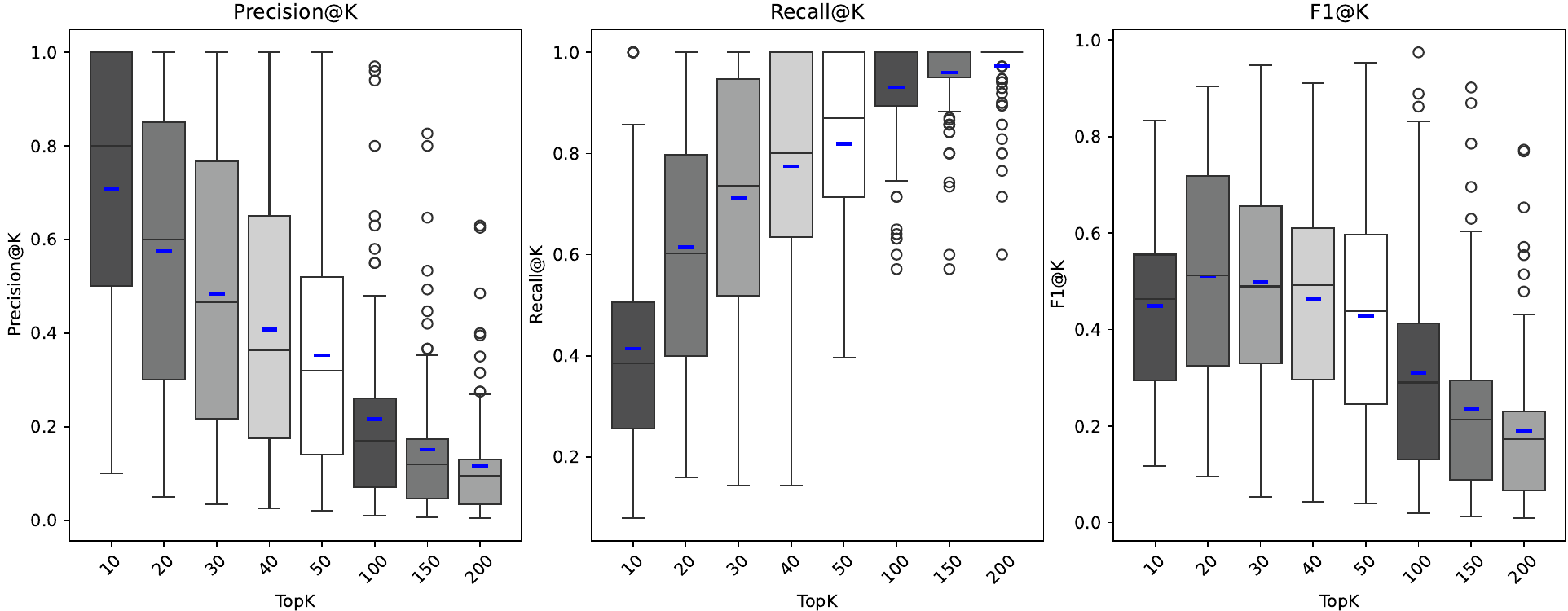}
% Precision-Recall Curve for Different K Values in Top-K
\caption{Performance measures of our approach over different $k$ Values in top-$k$.}\label{fig:sensitivityAnalysisK}
% \vspace{-0.3cm}
\end{figure}

Beyond the decision threshold, the choice of the top-$k$ predictions also plays a crucial role in ranking-based vulnerability identification.
The value of $k$ (i.e., the number of top-ranked CVE predictions to consider) directly impacts these metrics. Lower values of $k$ may decrease the number of false positives but increase the number of false negatives. 
selection of relevant CVEs for each attack description.
To identify the optimal value of $k$, we performed a sensitivity analysis using a balanced dataset from MITRE ATT\&CK composed of 50 positive and 50 negative instances of technique descriptions. For each value of $k$, we computed Precision and Recall and visualized their variation using boxplots. As shown in Fig.~\ref{fig:sensitivityAnalysisK}, the boxplot illustrates how the average Precision and Recall scores change as $k$ increases. At $k=20$, the average values of Precision and Recall converge, resulting in the highest F1 score. 
This indicates that $k=20$ offers the most balanced prediction performance, effectively guiding the selection of relevant CVEs for each attack description.

% \subsection{Model evaluation and validation}\label{sec:evaluation}

% To assess the performance of our model, we employed four distinct evaluation methods, each designed to examine a different aspect of similarity-based prediction accuracy. These methods include: (1) \textbf{M1 (Threshold-Based Detection)}, where a similarity score above a fixed threshold determines a match; (2) \textbf{M2 (Manual Detection)}, where the first author of this study validates the predictions through a manual inspection of the ground-truth attack-to-CVE associations; (3) \textbf{M3 (Similarity with First CVE)}, where only the first CVE description mentioned in the attack text is considered as the target for comparison; and (4)\textbf{ M4 (Similarity with All CVEs)}, where all CVEs mentioned in the attack description are taken into account for computing similarity scores.
\subsection{Model Evaluation and Validation}
\label{sec:evaluation}

To answer our RQs, we evaluate the effectiveness of our model in predicting software vulnerabilities from attack descriptions. The output of our model consists of a ranked list of top-$k$ vulnerability candidates for each attack description based on cosine similarity scores. 
We designed four evaluation strategies to validate the model's predictions: Three automated (oracle-based) methods and one manual. The methods target a distinct aspect of semantic similarity and practical performance.
Let $a$ denote an attack description, and let $C = \{c_1, c_2, ..., c_n\}$ be the set of vulnerability descriptions. Let $\text{sim}(a, c)$ represent the cosine similarity between the embeddings of $a$ and $c$, and let $\rho$ denote a fixed similarity threshold.

\paragraph{M1: Manual Detection.}
In this method, the study's two authors manually validated the predictions generated by the model. For each attack description, the top-$k$ CVEs predicted by the model were reviewed to determine whether they were semantically and contextually related to the attack text:

\[
\text{Predicted}_{\text{M1}}(a) = \{c_i \in C \mid c_i \text{ is judged related to } a \text{ by manual inspection}\}
\]

\paragraph{M2: Threshold-Based Detection.}
In this method, we consider as relevant all predicted vulnerabilities from the model’s top-$k$ list whose similarity score exceeds the threshold $\rho$. Since the optimal value of $\rho$ was already determined through our threshold-sensitivity analysis using a balanced subset of MITRE mappings (see Section~\ref{sec:sensitivityAnalysis}), we directly apply the resulting value $\rho = 0.58$ in this evaluation. This value corresponds to the Equal Error Rate (EER), where precision and recall intersect on the Precision–Recall curve (Fig.~\ref{fig:sensitivityAnalysis}). Using this calibrated threshold, the prediction set for M2 is defined as:
\[
\text{Predicted}_{\text{M2}}(a) = \{\,c \in \text{top-}k(C) \mid \text{sim}(a, c) \geq \rho\,\}.
\]

\paragraph{M3: Similarity with First CVE.}
This evaluation method investigates whether the model's predictions align semantically with the most vulnerable description mentioned in an attack news report. Instead of directly comparing the attack description to the CVE corpus, we extract the textual description of the first CVE explicitly referenced within the attack report. This CVE is assumed to represent the most central vulnerability discussed.

The embedding of this first CVE description, denoted as $c_1^{\text{attack}}$, is then compared to each predicted CVE from the model's output using cosine similarity. If the similarity score between the first CVE embedding and a predicted CVE exceeds the threshold $\rho$, the predicted CVE is considered a valid match:

\[
\text{Predicted}_{\text{M3}}(a) = \{c \in C \mid \text{sim}(c_1^{\text{attack}}, c) \geq \rho\}
\]

\paragraph{M4: Similarity with All CVEs.}
This evaluation method examines the model's ability to retrieve vulnerabilities semantically aligned with all CVEs mentioned in the attack report. Unlike M3, which uses a single CVE description, this method concatenates all CVE descriptions referenced in the attack report into a single text. The concatenated description is then embedded using the same sentence transformer model.

The resulting single embedding, which captures the combined semantics of all referenced vulnerabilities, is compared to the top-$k$ predictions generated by the model. If a predicted CVE has a similarity score above the threshold $\rho$ with the aggregated embedding, it is considered a correct prediction:

\[
\text{Predicted}_{\text{M4}}(a) = \{c \in \text{top-}k(C) \mid \text{sim}(\text{concat}(C')^{\text{attack}}, c) \geq \rho\}
\]

where $C' \subseteq C$ is the set of all CVEs explicitly mentioned in the attack report, and $\text{concat}(C')^{\text{attack}}$ represents the embedded concatenation of their descriptions.

These evaluation methods collectively validate the accuracy and reliability of the model’s predictions. By combining automated and manual assessment techniques, we ensure that the predicted vulnerabilities are quantitatively strong (high precision) and contextually meaningful about the original attack descriptions. This multi-validation process enhances the credibility of the evaluation and provides a rigorous foundation for assessing the model’s effectiveness in practical cybersecurity scenarios.

\section{Results and Discussion}
\label{sec:result}
In this section, we present the results of our experiments and the outcomes for the RQs.

\subsection{\rqone}
To answer RQ1, we evaluated the performance of our approach in predicting CVE issues from news descriptions about cyberattacks through manual validation (M1: Manual Detection) (see Section~\ref{sec:evaluation}).
% We evaluate our model on 100 SecurityWeek articles (see Section~\ref{sec:mapprocess} for dataset details).
\begin{table*}[htb]
\centering
\scriptsize
\caption{Validation outcomes for each method (M1–M4).}
\label{tab:cveComparison}
\begin{tabular}{lc|cc}
\hline
 & \textbf{Retrieved} & \multicolumn{2}{c}{\textbf{Validated}} \\ 
 & \textbf{Total} & \textbf{Relevant} & \textbf{Not Relevant} \\ 
\hline
M1: Manual Detection & 2000 & 1418 (70\%) & 582 (30\%) \\ 
M2: Threshold-Based Detection & 2000 & 1625 (81\%) & 375 (19\%) \\ 
M3: Similarity with First CVE & 2000 & 1607 (80\%) & 393 (20\%) \\ 
M4: Similarity with All CVEs & 2000 & 1575 (78\%) & 425 (22\%) \\ 
\hline
\end{tabular}
\end{table*}
We evaluate the model’s accuracy across 2000 predictions.
To assess their validity, two authors manually reviewed each predicted CVE by comparing its description to the content of the corresponding news report, determining whether the prediction was contextually and semantically relevant.

This manual validation served as the ground truth for evaluating model accuracy. Among the 2000 predicted CVEs, 1418 (70\%) were validated as relevant, demonstrating the model’s strong potential to support human analysts in identifying vulnerabilities from natural language threat reports. The detailed results of this validation process, including a comparison across all evaluation methods, are summarized in Table~\ref{tab:cveComparison}.
% \begin{table*}[htb]
% \centering
% \caption{Number of news reports where the model’s predicted CVEs matched at least one CVE mentioned in the report.}
% \label{tab:attackNewsCVE}
% \setlength{\tabcolsep}{5pt} % Adjust column spacing
% \renewcommand{\arraystretch}{1} % Adjust row spacing
% \resizebox{\textwidth}{!}{ % Resize to fit within page width
% \begin{tabular}{lcccc}
% \hline
%  & \multicolumn{4}{c}{\textbf{Retrieved}} \\ 
% \cline{2-5}
% \textbf{} & \textbf{Total} & \textbf{Attack News Containing Matching CVE-ID} & \textbf{Attack News No Matching CVE-ID} & \textbf{Attack News Without CVE-ID Link} \\ 
% \hline
% \textbf{Number of Attack News} & 100 & 57 (57\%) & 40 (40\%) & 3 (3\%) \\ 
% \hline
% \end{tabular}
% }
% \end{table*}
% Furthermore, as shown in Table~\ref{tab:attackNewsCVE}, we analyzed the overlap between the CVE identifiers explicitly mentioned in the attack news and those retrieved in the model's predictions. In 57\% of the cases, the predicted CVE list included at least one CVE ID that was also present in the corresponding news article, indicating strong alignment between the model's output and the ground-truth textual references. In contrast, 40\% of the news articles had no predicted CVE ID matching the referenced one, while 3\% of articles did not contain any CVE reference at all. This result further supports the validity of the model’s predictions and highlights its potential utility in practical threat intelligence applications.
Moreover, our model’s predictions overlapped with actual CVE mentions in 57 out of 100 news articles. In other words, the model predicts at least one exact CVE-ID truly mentioned, demonstrating a strong ability to pinpoint known vulnerabilities. The remaining cases include 40\% with only unseen (but possibly relevant) predictions and 3\% with no CVE mentions to match. Our approach can suggest plausible CVEs even when no explicit vulnerability identifiers (e.g., CVE-IDs) are mentioned in the news article, demonstrating its capability to infer hidden or implied vulnerabilities from unstructured text.
However, this result is influenced by the choice of the top-$k$ parameter, which was set to $k=20$ in our study. A larger $k$ could increase recall by including more potentially relevant CVEs in the prediction set, thereby improving the match rate. On the other hand, increasing $k$ may also reduce precision and introduce noise, which would require more effort in filtering or post-validation. In future work, we plan to validate the effect of varying $k$ values by comparing the model’s performance across multiple cybersecurity news and threat reports datasets. This comparison will allow us to identify patterns in prediction behavior and compute an average optimal $k$ value that balances precision and recall.

\subsection{\rqtwo}
To answer RQ2, we investigated the effectiveness of our automated oracle-based validation methods in assessing the semantic accuracy of CVE predictions generated by our approach from unstructured cyberattack news reports. These methods are designed to represent ground truth when manual annotation is unavailable or impractical. Our evaluation focuses on three distinct oracle methods: M2: Threshold-based detection,  M3: Similarity with First CVE, and M4: Similarity with All CVEs.

The M2 method evaluates model predictions by applying a similarity threshold $\rho$ to the top-$k$ ranked CVEs produced for each news report. By applying $\rho = 0.58$ to the model’s top-$k$ predictions, M2 yielded a precision of 81\%, indicating that the majority of CVEs retrieved at or above this similarity level were contextually relevant to the attack descriptions, as summarized in Table~\ref{tab:cveComparison}.

% The Similarity with First CVE (M3) strategy focuses on assessing how well the model’s predictions align with the core vulnerability discussed in the news article. In this method, we extract the description of the first CVE explicitly mentioned in the article and compute its semantic similarity with each of the predicted CVEs. This provides insight into whether the model captures the principal vulnerability referenced by the report. Using the calibrated threshold, this method achieved an accuracy of 80\%, highlighting the model’s capacity to retrieve semantically related vulnerabilities that align with the dominant threat mentioned.

The M3 method assesses whether the predicted vulnerabilities fall within the semantic proximity of the attack’s core vulnerability. Thus, we filtered the predictions using the similarity threshold ($\rho = 0.58$) to retain only those with a similarity above the threshold. Table~\ref{tab:cveComparison} shows that the M3 achieved an 80\% agreement with the first-mentioned CVE, suggesting that the model consistently retrieves CVEs that are semantically coherent with the primary vulnerability highlighted in the news report. These results demonstrate the model's capacity to capture subtle contextual relationships and reflect its potential utility in prioritizing the most relevant threats in real-world reporting.

The M4 method extends M3 by evaluating the semantic similarity between the model’s predicted CVEs and the combined set of all CVEs mentioned in the attack report. The results of M4 yielded a match rate of 78\%, indicating that the model can identify vulnerabilities that align with the overall threat context described in multi-CVE issues. This further highlights the model’s robustness in handling more complex and information-rich inputs, where multiple vulnerabilities contribute to the characterization of the attack.
The three automated validation methods (M2, M3, M4) all show roughly similar success rates (78–81\%), providing confidence that our model’s predictions align well with known vulnerabilities whether we compare them to a threshold or actual CVEs mentioned in reports.

\section{Threats to Validity}
\label{sec:threatstovalidity}
In this section, we illustrate the threats to the validity of our study.
\par\noindent
\textbf{Construct validity.}
Construct validity refers to the degree to which theoretical claims or hypotheses articulated at a conceptual level are supported by empirical evidence obtained from operational measures~\cite{sjoberg2022construct}. In the context of our study, construct validity explicitly addresses the efficacy of our evaluation in accurately capturing the concept of predicting pertinent vulnerabilities from attack descriptions. Two main threats may impact this. First, the cosine similarity threshold $\rho$ is not fixed; while we selected a value balancing false positives and false negatives, its tuning may vary by use case. A higher $\rho$ favors precision, while a lower one favors recall. Second, in M3 and M4, we assume that the first-mentioned or aggregated CVEs in a report reflect ground truth. However, some reports may reference irrelevant or no CVEs. We mitigate these risks through consistent thresholds and leveraging real-world datasets from MITRE. 
\par\noindent
\textbf{Internal validity.}
% Internal validity refers to the extent to which the outcomes of our study can be confidently attributed to the proposed model and evaluation procedures. One potential threat is the selection of the dataset. 
% Although we used 100 news reports from a reputable source (SecurityWeek), this may not fully capture the variety of writing styles and reporting structures found in cybersecurity news. Additionally, the language used in these reports can vary considerably; some contain highly structured technical content, including version numbers or embedded CVE-IDs, which often deviate from conventional natural language expressions.
% Moreover, the use of a fixed top-$k$ value for all predictions is employed. While using $k=20$ allows for a consistent comparison, it may not be equally optimal for all cases, potentially leading to underestimation of the model’s performance. 
% Despite these limitations, we addressed internal consistency by applying uniform preprocessing steps, using a similarity threshold, and validating across multiple independent evaluation methods.
Internal validity reflects how confidently the results can be attributed to our model and evaluation design~\cite{wohlin2012experimentation}. A potential threat lies in dataset selection; while the 100 SecurityWeek reports are credible, they may not reflect the full diversity of writing styles in cybersecurity reporting. Some texts include structured technical content or CVE references that differ from natural language patterns. Additionally, using a fixed top-$k$ value ($k=20$) ensures consistency but may not suit all cases equally. To mitigate these concerns, we applied uniform pre-processing, used a similarity threshold, and validated performance across multiple independent methods.
\par\noindent
\textbf{External validity.}
% External validity refers to the extent to which the findings of this study can be generalized beyond the specific dataset and experimental setup. A potential threat arises from the fact that our evaluation was conducted using a dataset of 100 cybersecurity news reports exclusively collected from SecurityWeek. 
% Although this source is credible and widely cited, relying on a single outlet may limit the generalizability of the results, as different news providers may present attack information with varying levels of detail, structure, and terminology. Furthermore, our method was tested only on English-language reports and may not directly generalize to multilingual contexts or non-news sources such as social media, blogs, or technical advisories. To improve external validity, future work should validate the approach on more diverse datasets and across different types of cyber threat intelligence sources.
External validity concerns the generalizability of our findings beyond the current dataset and setup~\cite{wohlin2012experimentation}. Our evaluation used 100 cybersecurity news reports from SecurityWeek, a reputable but single source, which may limit applicability to other news sources with different styles and terminology. Additionally, the method was tested only on English-language reports, excluding multilingual or alternative sources like social media or technical blogs. 
Additionally, our evaluation set of 100 news reports can be considered limited; expanding this number in future work would increase confidence in the generalizability of the results and validate the approach on broader and more diverse datasets. Additionally, our approach uses only the MPNet model, which may limit generalizability. In future work, we aim to test other transformer models and domain-adapted variants to improve robustness.

\section{Related Work}
\label{ch:relatedwork}
In this section, we provide an overview of related work on vulnerability-attack models. This section summarizes recent advancements in this domain and contrasts them with our proposed approach.
Kuppa et al.~\cite{kuppa2021linking} proposed a multi-head deep embedding model to link CVE issues with MITRE ATT\&CK techniques. Their method involved extracting relevant information from CVE metadata using regular expressions and comparing it with ATT\&CK vectors via cosine similarity. However, this approach was limited to a small subset of ATT\&CK techniques. Similarly, Sun et al.~\cite{sun2021generating} employed a BERT-based model~\cite{devlin2018bert} to enrich textual CVE descriptions, aiding downstream information extraction and linkage tasks.
Other studies approached the problem from a multi-label classification perspective. Lakhdhar et al.~\cite{lakhdhar2021machine} experimented with various traditional and deep learning algorithms to map CVEs to ATT\&CK tactics. Grigorescu et al.~\cite{grigorescu2022cve2att} introduced CVE2ATT\&CK, a model that annotated CVE issues with relevant ATT\&CK tactics using both BERT-based and classical machine learning models. Ampel et al.~\cite{ampel2021linking} proposed CVE Transformer (CVET), which incorporates a self-distillation mechanism for fine-tuning RoBERTa~\cite{liu2019roberta} to associate CVEs with one of ten ATT\&CK tactics.
Several studies have focused on linking attack patterns (e.g., CAPEC) to vulnerabilities (CVE) rather than leveraging Tactics, Techniques, and Procedures (TTPs). Kanakogi et al.~\cite{kanakogi2021tracing,kanakogi2022comparative} and Hemberg et al.~\cite{hemberg2022sourcing} used NLP techniques, including RoBERTa, to compute semantic similarity between CAPEC and CVE descriptions. TF-IDF~\cite{ramos2003using} and Doc2Vec~\cite{lau2016empirical} were also evaluated, with TF-IDF showing the most effective matching in ranking top-N relevant CAPEC documents for each CVE.

Prior work has explored reversing the typical mapping direction by predicting CVEs from textual descriptions of ATT\&CK techniques~\cite{othman2024cybersecurity}. One study introduced an automated tool leveraging nine sentence transformer models to perform this mapping, offering fine-grained linkage based on real-world procedures. Another comparative study~\cite{othman2024comparison} evaluated five feature extraction techniques (TF-IDF, LSI, BERT, MiniLM, RoBERTa) for linking CAPEC attack patterns to CVEs, resulting in a comprehensive mapping dataset that connects 133 CAPEC patterns to 685 CVEs through shared weaknesses.
%for review
% Our prior work extends these efforts by reversing the typical mapping direction. In VULDAT~\cite{othman2024cybersecurity}, we proposed an automated tool leveraging nine sentence transformer models to predict CVEs from textual descriptions of ATT\&CK techniques. Unlike CAPEC-based studies, VULDAT focuses exclusively on mapping attack technique descriptions to CVEs, providing a higher granularity of linkage based on real-world attack procedures.
% Additionally, in our follow-up study~\cite{othman2024comparison}, we conducted an empirical comparison of feature extraction techniques (TF-IDF, LSI, BERT, MiniLM, RoBERTa) to evaluate their effectiveness in linking CAPEC attack patterns to CVEs. The study resulted in a comprehensive mapping dataset that connects 133 CAPEC patterns to 685 CVEs through shared weaknesses.

%for review
% While most studies focus on enhancing CVE data or linking known vulnerabilities to attacks using structured sources like MITRE’s CVE, CAPEC, or ATT\&CK, our work takes a different approach: predicting CVEs directly from real-world cyberattack news. This is a challenging task due to the unstructured nature of news text and the lack of explicit technical details. To our knowledge, this is the first study to automatically link attack news reports to CVE-IDs using sentence transformers, offering a novel way to enhance vulnerability awareness and response.

While most studies focus on enhancing CVE data or linking known vulnerabilities to attacks using structured sources such as MITRE’s CVE, CAPEC, or ATT\&CK, a different approach has been proposed: predicting CVEs directly from real-world cyberattack news. This task is particularly challenging due to the unstructured nature of news text and the frequent absence of explicit technical details. Recent work explores the automatic linking of attack news reports to CVE-IDs using sentence transformer models, offering a novel direction to improve vulnerability awareness and incident response.

\section{Conclusion and Future Work}
\label{ch:conclusion} 
In this study, we introduced an innovative approach for predicting software vulnerabilities directly from unstructured cyberattack news reports using the MPNet sentence transformer model. Our proposed method generates a ranked list of top-K CVE predictions based on the semantic similarity between attack descriptions and vulnerability reports. To ensure a robust evaluation, we developed four validation strategies: threshold-based detection, manual expert review, similarity with the first mentioned CVE, and similarity with all mentioned CVEs. Our results highlight the model's effectiveness, achieving validation accuracy of up to 81\% depending on the strategy employed, confirming the practical feasibility of mapping attack narratives to vulnerability databases such as the CVE repository.
Additionally, we assessed our model using a dataset of 100 cybersecurity news articles, analyzing each to determine the relevance of the predicted CVEs. This evaluation methodology, which combined threshold analysis, semantic similarity, and manual validation, facilitated a comprehensive assessment of the model's predictive accuracy. Our approach quantified performance and revealed the model's strengths in identifying relevant vulnerabilities from unstructured textual sources.

For future research, we intend to explore the influence of the top-$K$ parameter on prediction accuracy. While this study utilized a fixed value (e.g., $k=20$), employing a dynamic or optimized $K$-value could enhance the balance between precision and recall, especially across various attack types and news formats. Additionally, our current evaluation is based exclusively on reports from SecurityWeek. To improve generalizability, we plan to expand our evaluation to include additional datasets and test our approach on varied sources beyond SecurityWeek. Moreover, we aim to establish connections between attack reports and their corresponding vulnerability codes (e.g., CWE IDs). This approach would facilitate more accurate and explainable linkages between textual descriptions of attacks and formal vulnerability databases, ultimately enhancing the validation and traceability of predicted CVEs.

\section*{Acknowledgements}
\begin{sloppypar}
This research was supported by the European Social Fund Plus (ESF+), Project ESF2f30005, CUP: B56F24000100001. The authors also gratefully acknowledge the support of the Cybersecurity Laboratory (CSLab) at the Free University of Bozen-Bolzano funded by the EFRE-FESR 2021–2027 program, project EFRE1039, CUP: I53C23001690009.
\end{sloppypar}

\printbibliography[heading=bibintoc]

@misc{NVDvulnerability,
      author = "NVD",
       title = "NVD Vulnerabilities",
	note = "\url{https://nvd.nist.gov/vuln}"
}

@misc{CResearch,
      author = "Check Point",
       title = "38\% Increase in 2022 Global Cyberattacks",
	note = "\url{https://blog.checkpoint.com/2023/01/05/38-increase-in-2022-global-cyberattacks/}"
}

@misc{2024Report,
      author = "Check Point",
       title = "Shifting Attack Landscapes and Sectors in Q1 2024 with a 28\% increase in cyber attacks globally",
	note ="\url{https://blog.checkpoint.com/research/shifting-attack-landscapes-and-sectors-in-q1-2024-with-a-28-increase-in-cyber-attacks-globally/}"
}

@article{elder2022really,
  title={Do I really need all this work to find vulnerabilities? An empirical case study comparing vulnerability detection techniques on a Java application},
  author={Elder, Sarah and Zahan, Nusrat and Shu, Rui and Metro, Monica and Kozarev, Valeri and Menzies, Tim and Williams, Laurie},
  journal={Empirical Software Engineering},
  volume={27},
  number={6},
  pages={154},
  year={2022},
  publisher={Springer}
}

@misc{multi-qa-mpnet-base-dot-v1,
   author={Hugging Face },  title = "multi-qa-mpnet-base-dot-v1",
  year = "2024",
  note = "Accessed: May 2, 2024. \url{https://huggingface.co/sentence-transformers/multi-qa-mpnet-base-dot-v1}"
}

@inproceedings{kuppa2021linking,
  title={Linking cve’s to mitre att\&ck techniques},
  author={Kuppa, Aditya and Aouad, Lamine and Le-Khac, Nhien-An},
  booktitle={Proceedings of the 16th International Conference on Availability, Reliability and Security},
  pages={1--12},
  year={2021}
}

@article{song2020mpnet,
  title={Mpnet: Masked and permuted pre-training for language understanding},
  author={Song, Kaitao and Tan, Xu and Qin, Tao and Lu, Jianfeng and Liu, Tie-Yan},
  journal={Advances in neural information processing systems},
  volume={33},
  pages={16857--16867},
  year={2020}
}

@inproceedings{othman2024comparison,
  title={A comparison of vulnerability feature extraction methods from textual attack patterns},
  author={Othman, Refat and Rossi, Bruno and Russo, Barbara},
  booktitle={2024 50th Euromicro Conference on Software Engineering and Advanced Applications (SEAA)},
  pages={419--422},
  year={2024},
  organization={IEEE}
}

@inproceedings{othman2024cybersecurity,
  title={Cybersecurity defenses: Exploration of cve types through attack descriptions},
  author={Othman, Refat and Rossi, Bruno and Russo, Barbara},
  booktitle={2024 50th Euromicro Conference on Software Engineering and Advanced Applications (SEAA)},
  pages={415--418},
  year={2024},
  organization={IEEE}
}

@misc{VULDAT,
    author={Refat Othman},
  title = {ATT\&CK2VUL - Automated Vulnerability Detection From Cyberattack Text},
  year = {2025},
  note = {Accessed: Feb 2, 2025. \url{https://github.com/ref3t/Attack2VUL/tree/main}}
}

@misc{CVEdataset,
      author = "MITRE",
       title = "CVE",
	note = "\url{https://cve.mitre.org/}"
}

@misc{CWE,
      author = "MITRE",
       title = "CWE Dataset",
	note = "\url{https://cwe.mitre.org/}"
}

@misc{CAPEC,
      author = "MITRE",
       title = "CAPEC",
	note = "\url{https://capec.mitre.org/}"
}

@misc{ATTACK,
      author = "MITRE",
       title = "ATTACK",
	note = "\url{https://attack.mitre.org/}"
}

@misc{CVSS,
      author = "NVD",
       title = "CVSS",
	note = "\url{https://nvd.nist.gov/vuln-metrics/cvss/v3-calculator}"
}

@misc{CTI,
  author = "Gartner",
  title = "Threat Intelligence",
  note = "\url{https://www.gartner.com/en/documents/2487216}"
}

@misc{WhatCVE,
  author = "Taylor Armerding",
  title = "CVE Definitions",
  note = "\url{https://www.csoonline.com/article/3204884/what-is-cve-its-definition-and-purpose.html}"
}

@phdthesis{baccar2021automated,
  title={Automated mapping of CVE vulnerabilties to MITRE ATT\&CK Framework},
  author={Baccar, Karim},
  year={2021},
  school={Tekup}
}

@misc{CTIINfo,
  title = "What is Threat Intelligence?",
  author={Staff Contributor},
  note = "\url{https://www.dnsstuff.com/what-is-threatintelligence}"
}

@article{gasmi2019information,
  title={Information extraction of cybersecurity concepts: An LSTM approach},
  author={Gasmi, Houssem and Laval, Jannik and Bouras, Abdelaziz},
  journal={Applied Sciences},
  volume={9},
  number={19},
  pages={3945},
  year={2019},
  publisher={MDPI}
}

@inproceedings{alevizopoulou2021social,
  title={Social media monitoring for IoT cyber-threats},
  author={Alevizopoulou, Sofia and Koloveas, Paris and Tryfonopoulos, Christos and Raftopoulou, Paraskevi},
  booktitle={2021 IEEE International Conference on Cyber Security and Resilience (CSR)},
  pages={436--441},
  year={2021},
  organization={IEEE}
}

@inproceedings{iorga2021yggdrasil,
  title={Yggdrasil—early detection of cybernetic vulnerabilities from Twitter},
  author={Iorga, Denis and Corlatescu, Dragos-Georgian and Grigorescu, Octavian and Sandescu, Cristian and Dascalu, Mihai and Rughinis, Razvan},
  booktitle={2021 23rd International Conference on Control Systems and Computer Science (CSCS)},
  pages={463--468},
  year={2021},
  organization={IEEE}
}

@inproceedings{queiroz2019eavesdropping,
  title={Eavesdropping hackers: Detecting software vulnerability communication on social media using text mining},
  author={Queiroz, Andrei Lima and Mckeever, Susan and Keegan, Brian},
  booktitle={The Fourth International Conference on Cyber-Technologies and Cyber-Systems},
  pages={41--48},
  year={2019}
}

@inproceedings{dionisio2019cyberthreat,
  title={Cyberthreat detection from twitter using deep neural networks},
  author={Dion{\'\i}sio, Nuno and Alves, Fernando and Ferreira, Pedro M and Bessani, Alysson},
  booktitle={2019 international joint conference on neural networks (IJCNN)},
  pages={1--8},
  year={2019},
  organization={IEEE}
}

@article{kanakogi2021tracing,
  title={Tracing cve vulnerability information to capec attack patterns using natural language processing techniques},
  author={Kanakogi, Kenta and Washizaki, Hironori and Fukazawa, Yoshiaki and Ogata, Shinpei and Okubo, Takao and Kato, Takehisa and Kanuka, Hideyuki and Hazeyama, Atsuo and Yoshioka, Nobukazu},
  journal={Information},
  volume={12},
  number={8},
  pages={298},
  year={2021},
  publisher={MDPI}
}

@inproceedings{ramos2003using,
  title={Using tf-idf to determine word relevance in document queries},
  author={Ramos, Juan and others},
  booktitle={Proceedings of the first instructional conference on machine learning},
  volume={242},
  number={1},
  pages={29--48},
  year={2003},
  organization={Citeseer}
}

@article{lonetti2023model,
  title={Model-based security testing in IoT systems: A Rapid Review},
  author={Lonetti, Francesca and Bertolino, Antonia and Di Giandomenico, Felicita},
  journal={Information and Software Technology},
  pages={107326},
  year={2023},
  publisher={Elsevier}
}

@article{liu2022context2vector,
  title={Context2Vector: Accelerating security event triage via context representation learning},
  author={Liu, Jia and Zhang, Runzi and Liu, Wenmao and Zhang, Yinghua and Gu, Dujuan and Tong, Mingkai and Wang, Xingkai and Xue, Jianxin and Wang, Huanran},
  journal={Information and Software Technology},
  volume={146},
  pages={106856},
  year={2022},
  publisher={Elsevier}
}

@article{lau2016empirical,
  title={An empirical evaluation of doc2vec with practical insights into document embedding generation},
  author={Lau, Jey Han and Baldwin, Timothy},
  journal={arXiv preprint arXiv:1607.05368},
  year={2016}
}

@inproceedings{lakhdhar2021machine,
  title={Machine learning based approach for the automated mapping of discovered vulnerabilities to adversial tactics},
  author={Lakhdhar, Yosra and Rekhis, Slim},
  booktitle={2021 IEEE Security and Privacy Workshops (SPW)},
  pages={309--317},
  year={2021},
  organization={IEEE}
}

@article{kanakogi2022comparative,
  title={Comparative Evaluation of NLP-Based Approaches for Linking CAPEC Attack Patterns from CVE Vulnerability Information},
  author={Kanakogi, Kenta and Washizaki, Hironori and Fukazawa, Yoshiaki and Ogata, Shinpei and Okubo, Takao and Kato, Takehisa and Kanuka, Hideyuki and Hazeyama, Atsuo and Yoshioka, Nobukazu},
  journal={Applied Sciences},
  volume={12},
  number={7},
  pages={3400},
  year={2022},
  publisher={MDPI}
}

@article{devlin2018bert,
  title={Bert: Pre-training of deep bidirectional transformers for language understanding},
  author={Devlin, Jacob and Chang, Ming-Wei and Lee, Kenton and Toutanova, Kristina},
  journal={arXiv preprint arXiv:1810.04805},
  year={2018}
}

@article{grigorescu2022cve2att,
  title={Cve2att\&ck: Bert-based mapping of cves to mitre att\&ck techniques},
  author={Grigorescu, Octavian and Nica, Andreea and Dascalu, Mihai and Rughinis, Razvan},
  journal={Algorithms},
  volume={15},
  number={9},
  pages={314},
  year={2022},
  publisher={MDPI}
}

@article{sun2021generating,
  title={Generating informative CVE description from ExploitDB posts by extractive summarization},
  author={Sun, Jiamou and Xing, Zhenchang and Guo, Hao and Ye, Deheng and Li, Xiaohong and Xu, Xiwei and Zhu, Liming},
  journal={arXiv preprint arXiv:2101.01431},
  year={2021}
}

@article{ampel2021linking,
  title={Linking common vulnerabilities and exposures to the mitre att\&ck framework: A self-distillation approach},
  author={Ampel, Benjamin and Samtani, Sagar and Ullman, Steven and Chen, Hsinchun},
  journal={arXiv preprint arXiv:2108.01696},
  year={2021}
}

@article{hemberg2022sourcing,
  title={Sourcing Language Models and Text Information for Inferring Cyber Threat, Vulnerability and Mitigation Relationships},
  author={Hemberg, Erik and Srinivasan, Ashwin and Rutar, Nick and O’Reilly, Una-May},
  year={2022}
}

@article{liu2019roberta,
  title={Roberta: A robustly optimized bert pretraining approach},
  author={Liu, Yinhan and Ott, Myle and Goyal, Naman and Du, Jingfei and Joshi, Mandar and Chen, Danqi and Levy, Omer and Lewis, Mike and Zettlemoyer, Luke and Stoyanov, Veselin},
  journal={arXiv preprint arXiv:1907.11692},
  year={2019}
}

@inproceedings{vuldatPaper222,
  title={VULDAT: Automated Vulnerability Detection From Cyberattack Text},
  author={Othman, Refat and Russo, Barbara},
  booktitle={Embedded Computer Systems: Architectures, Modeling, and Simulation: 23rd International Conference, SAMOS},
  year={2023}
}

@inproceedings{catillo2021critique,
  title={A critique on the use of machine learning on public datasets for intrusion detection},
  author={Catillo, Marta and Del Vecchio, Andrea and Pecchia, Antonio and Villano, Umberto},
  booktitle={Quality of Information and Communications Technology: 14th International Conference, QUATIC 2021, Algarve, Portugal, September 8--11, 2021, Proceedings 14},
  pages={253--266},
  year={2021},
  organization={Springer}
}

@inproceedings{davis2006relationship,
  title={The relationship between Precision-Recall and ROC curves},
  author={Davis, Jesse and Goadrich, Mark},
  booktitle={Proceedings of the 23rd international conference on Machine learning},
  pages={233--240},
  year={2006}
}

@inproceedings{sheng2006thresholding,
  title={Thresholding for making classifiers cost-sensitive},
  author={Sheng, Victor S and Ling, Charles X},
  booktitle={Aaai},
  volume={6},
  pages={476--481},
  year={2006}
}

@inproceedings{lobo2022cost,
  title={Cost-Sensitive Learning and Threshold-Moving Approach to Improve Industrial Lots Release Process on Imbalanced Datasets},
  author={Lobo, Armindo and Oliveira, Pedro and Sampaio, Paulo and Novais, Paulo},
  booktitle={International Symposium on Distributed Computing and Artificial Intelligence},
  pages={280--290},
  year={2022},
  organization={Springer}
}

@inproceedings{othman2024vulnerability,
  title={Vulnerability Detection for software-intensive system},
  author={Othman, Refat Tahseen},
  booktitle={Proceedings of the 28th International Conference on Evaluation and Assessment in Software Engineering},
  pages={510--515},
  year={2024}
}

@article{HinidumaEtAl2025, 
author = {Hiniduma, Kaveen and Byna, Suren and Bez, Jean Luca}, title = {Data Readiness for AI: A 360-Degree Survey}, year = {2025}, issue_date = {September 2025}, publisher = {Association for Computing Machinery}, address = {New York, NY, USA}, volume = {57}, number = {9}, issn = {0360-0300}, url = {https://doi.org/10.1145/3722214}, doi = {10.1145/3722214},  journal = {ACM Comput. Surv.}, month = apr, articleno = {219}, numpages = {39}}

@misc{hyperparams,
  title = "Semantic Textual Similarity",
  note = "\url{https://github.com/UKPLab/sentence-transformers/blob/master/examples/sentence_transformer/training/sts/README.md}"
}

@article{muennighoff2022mteb,
  title={Mteb: Massive text embedding benchmark},
  author={Muennighoff, Niklas and Tazi, Nouamane and Magne, Lo{\"\i}c and Reimers, Nils},
  journal={arXiv preprint arXiv:2210.07316},
  year={2022}
}

@inproceedings{reimers2019sentence,
  title={Sentence-BERT: Sentence Embeddings using Siamese BERT-Networks},
  author={Reimers, Nils and Gurevych, Iryna},
  booktitle={Proceedings of the 2019 Conference on Empirical Methods in Natural Language Processing and the 9th International Joint Conference on Natural Language Processing (EMNLP-IJCNLP)},
  year={2019},
  organization={Association for Computational Linguistics}
}

@article{siino2024text,
  title={Is text preprocessing still worth the time? A comparative survey on the influence of popular preprocessing methods on Transformers and traditional classifiers},
  author={Siino, Marco and Tinnirello, Ilenia and La Cascia, Marco},
  journal={Information Systems},
  volume={121},
  pages={102342},
  year={2024},
  publisher={Elsevier}
}

@inproceedings{okonkwo2023leveraging,
  title={Leveraging word embeddings and transformers to extract semantics from building regulations text},
  author={Okonkwo, Odinakachukwu and Dridi, Amna and Vakaj, Edlira},
  booktitle={Proceedings of the 11th Linked Data in Architecture and Construction Workshop},
  year={2023}
}

@article{othman2025attack,
  title={From attack descriptions to vulnerabilities: A sentence transformer-based approach},
  author={Othman, Refat and Rimawi, Diaeddin and Rossi, Bruno and Russo, Barbara},
  journal={Journal of Systems and Software},
  pages={112615},
  year={2025},
  publisher={Elsevier}
}

@article{tang2023csgvd,
  title={CSGVD: A deep learning approach combining sequence and graph embedding for source code vulnerability detection},
  author={Tang, Wei and Tang, Mingwei and Ban, Minchao and Zhao, Ziguo and Feng, Mingjun},
  journal={Journal of Systems and Software},
  volume={199},
  pages={111623},
  year={2023},
  publisher={Elsevier}
}

@misc{VULDATDataSet,
    author={Refat Othman},
  title = {VULDAT- Vulnerability DataSet},
  year = {2025},
  note = {Accessed: Feb 2, 2025. \url{  figshare. Dataset. https://doi.org/10.6084/m9.figshare.25828102.v1}}
}

@inproceedings{husari2017ttpdrill,
  title={Ttpdrill: Automatic and accurate extraction of threat actions from unstructured text of cti sources},
  author={Husari, Ghaith and Al-Shaer, Ehab and Ahmed, Mohiuddin and Chu, Bill and Niu, Xi},
  booktitle={Proceedings of the 33rd annual computer security applications conference},
  pages={103--115},
  year={2017}
}

@article{jo2022vulcan,
  title={Vulcan: Automatic extraction and analysis of cyber threat intelligence from unstructured text},
  author={Jo, Hyeonseong and Lee, Yongjae and Shin, Seungwon},
  journal={Computers \& Security},
  volume={120},
  pages={102763},
  year={2022},
  publisher={Elsevier}
}

@article{wang2024knowcti,
  title={KnowCTI: Knowledge-based cyber threat intelligence entity and relation extraction},
  author={Wang, Gaosheng and Liu, Peipei and Huang, Jintao and Bin, Haoyu and Wang, Xi and Zhu, Hongsong},
  journal={Computers \& Security},
  volume={141},
  pages={103824},
  year={2024},
  publisher={Elsevier}
}

@misc{SecurityWeek,
      author = "SecurityWeek",
       title = "News Vulnerabilities",
	note = "\url{https://www.securityweek.com/category/vulnerabilities/}"
}

@article{sjoberg2022construct,
  title={Construct validity in software engineering},
  author={Sj{\o}berg, Dag IK and Bergersen, Gunnar Rye},
  journal={IEEE Transactions on Software Engineering},
  volume={49},
  number={3},
  pages={1374--1396},
  year={2022},
  publisher={IEEE}
}

@book{wohlin2012experimentation,
  title={Experimentation in software engineering},
  author={Wohlin, Claes and Runeson, Per and H{\"o}st, Martin and Ohlsson, Magnus C and Regnell, Bj{\"o}rn and Wessl{\'e}n, Anders and others},
  volume={236},
  year={2012},
  publisher={Springer}
}

@inproceedings{regano2024privacy,
  title={A Privacy-Preserving Approach for Vulnerability Scanning Detection},
  author={Regano, Leonardo and Canavese, Daniele and Mannella, Luca},
  booktitle={ITASEC 2024: 8th Italian Conference on Cyber Security},
  year={2024}
}

\end{document}